\documentclass[11pt]{article}
    
\usepackage{amsmath,amsfonts,amsthm,amscd,amssymb,graphicx}
\usepackage{xcolor} \usepackage{graphicx}

\numberwithin{equation}{section}
\usepackage[hidelinks]{hyperref}

\newtheorem{theorem}{Theorem}[section]

\newtheorem{lemma}[theorem]{Lemma}

\newtheorem{proposition}[theorem]{Proposition}

\newtheorem{remark}[theorem]{Remark}

\newcommand{\RR}{\mathbb{R}}

\newcommand{\cZ}{\mathcal{Z}}

\newcommand{\cH}{\mathcal{H}}

\newcommand{\PP}{\mathbb{P}}

\DeclareMathOperator{\Tr}{Tr}
\newcommand{\cD}{\mathcal{D}}
\newcommand{\cT}{\mathcal{T}}
\newcommand{\cE}{\mathcal{E}}
\newcommand{\aeff}{a_{\mathrm{eff}}}

\newcommand{\vrho}{\varrho}
\newcommand{\vrholo}{\underline\vrho}
\newcommand{\Lam}{\Lambda}

\newcommand{\gam}{\gamma}
\newcommand{\om}{\omega}

\newcommand{\onwl}{\operatornamewithlimits}
\newcommand{\diy}{\displaystyle}

\newcommand{\rL}{\mathrm{L}}

\allowdisplaybreaks[2]

\begin{document}

\title{A refined upper bound on the critical temperature for dilute Bose gases}

\author{
	P. T. Nam\thanks{LMU Munich, Department of Mathematics,
		Theresienstr.~39, 80333 Munich, Germany.
		Email: \texttt{nam@math.lmu.de}. 		}
	\and
	T. T. Nguyen\thanks{Penn State University, Department of Mathematics,
		State College, PA 16802, USA.
		Emails: \texttt{nguyen@math.psu.edu},
		\texttt{ius68@psu.edu}, \texttt{ims14@psu.edu}.
}
	\and
	I. Stuhl\footnotemark[2]
	\and
	Y. Suhov\footnotemark[2]
}

%\date{September 24, 2026}

\maketitle

\begin{abstract}
We establish an upper bound on the critical temperature shift of three-dimensional Bose gases with general nonnegative finite-range interactions. In the dilute regime $a\rho^{1/3}\to0$, we prove $T_c(\rho)/T_c^{(0)}(\rho) \le 1 + (C_{\rm H}+o(1))\sqrt{a\rho^{1/3}}$, where
\[
 C_{\rm H}=\frac{2^{7/2}\sqrt\pi}{3\zeta(3/2)^{2/3}}
 =3.523992\ldots
\]
is the coefficient predicted by Huang in 1999. This improves the earlier rigorous upper bound of Seiringer and Ueltschi in 2009. More precisely, we prove the exponential decay of the correlations above the stated temperature through the Feynman--Kac representation, for both integrable and hard-core interactions. 
\end{abstract}

%%%%%%%%%%%%%%%%%%%%%
\section{Introduction}\label{sec-intro}
%%%%%%%%%%%%%%%%%%%%%

We are interested in the critical temperature at which Bose--Einstein condensation (BEC) occurs in an interacting Bose gas. The ideal-gas transition is explicit, whereas in the interacting problem two-body scattering is coupled to long-wavelength critical fluctuations. 
For an interacting system, the Penrose--Onsager criterion identifies
condensation with a macroscopic eigenvalue of the one-particle density
matrix~\cite{PO}. Our aim is to bound the temperature above which this cannot
occur. We obtain a sharper normal-phase bound by refining the path-integral
method of Seiringer and Ueltschi~\cite{SU}.

\subsection{The setting}\label{sec-model}

Throughout the paper, $\Lam\subset\RR^3$ is a bounded open  
domain with a piecewise smooth boundary, and $U:x\in\RR^3
\mapsto [0,\infty]$ is an even function, supported in a ball of radius 
$R_0<\infty$ and continuous at each point $x$ where $U(x)<\infty$. For every integer $N\geq 2$, we consider the $N$-particle bosonic Hamiltonian with the two-body potential $U$:
\begin{equation}\label{HN}
 H_{\Lambda,N}=\sum_{i=1}^N(-\Delta_i)
       +\sum_{1\le i<j\le N}U(x_i-x_j)
\end{equation}
acting on $L^2(\Lambda)^{\otimes_sN}$, with Dirichlet boundary conditions. For $N=1$, the potential term is absent and  $H_{\Lambda,1}= -\Delta$. Singular repulsions and hard cores are understood by monotone approximation
with $U_\ell=\min\{U,\ell\}$ and $\ell\to \infty$ (in particular, this implies the Dirichlet boundary
condition on the hard core hypersurface). All semigroups and path measures below have
this interpretation. For a hard core, the semigroup is extended by
zero on excluded configurations. Our units are $\hbar=2m=k_B=1$. The case when $U=0$ is referred to as an ideal Bose gas.

At temperature $T=\beta^{-1}$, the grand-canonical partition function and particle density in volume $\Lambda$ are defined by 
\begin{equation}\label{canZ}
 \cZ_\Lambda(\beta,z)
   =1+\sum_{N\ge 1}z^N\Tr(e^{-\beta H_{\Lambda,N}})
\end{equation}
and
\begin{equation}\label{density}
 \vrho_{\Lambda,\beta}(z)
 =\frac{z}{|\Lambda|}\partial_z\log\cZ_\Lambda(\beta,z)
 =\frac{1}{|\Lambda|\cZ_\Lambda(\beta,z)}
   \sum_{N\ge 1}Nz^N\Tr(e^{-\beta H_{\Lambda,N}})
=:\dfrac{1}{|\Lam |}{\mathbb E}_{\Lam,\beta ,z}[N].
\end{equation}
Here the parameter $z>0$ represents the fugacity, and the parameter $\mu =\beta^{-1}\ln z$ represents the chemical
potential. Next, ${\mathbb E}_{\Lam,\beta ,z}[N]=\dfrac{1}{\cZ_\Lambda(\beta,z)}
 \sum\limits_{N\ge 1}Nz^N\Tr(e^{-\beta H_{\Lambda,N}})$ 
stands for the expected particle number in the grand-canonical Gibbs state. The one-particle (reduced) density matrix kernel in $\Lam$ is defined by 
\begin{equation}\label{1prdm}
\gamma_{\Lam,\beta}(x,y;z)= \frac{1}{\cZ_{\Lam,\beta}(z)}\sum\limits_{N\ge 1}z^N\Tr(e^{-\beta H_{\Lam,N}}a^*(x)a(y)),\quad x,y\in\Lam 
\end{equation}
where $a^*(x),a(y)$ are the standard bosonic creation and annihilation operator-valued
distributions with the canonical commutation relations 
\begin{equation}\label{CCR}
[a(x), a(y)]=0=[a^*(x),a^*(y)], \;\; [a(x),a^*(y)]=\delta(x-y),\quad x,y\in{\mathbb R}^3. 
\end{equation}

For $0<z<1$, the partition function and the numerator series in \eqref{density} and 
\eqref{1prdm} are dominated by their ideal-gas
counterparts and are finite. We retain the dependence on $\Lam$ 
and $\beta$ when it is needed, and otherwise write  $\cZ(z)$, $\vrho(z)$ 
and $\gamma(x,y;z)$, referring to \eqref{canZ}, \eqref{density} and \eqref{1prdm}.

We are interested in the  thermodynamic limit $\Lam=\Lam_L\uparrow\RR^3$ 
with $\Lam_L=[-L/2,L/2]^3$ and $L\to\infty$, where $z=z(L ,\beta)$
varies with $L$ in such a way that $\vrho_{\Lam_L,\beta}(z)\to\rho$. 
The critical temperature $T_c (\rho )$ is defined as the upper endpoint of the set of temperatures at which off-diagonal long-range order occurs in the thermodynamic limit. 
 Formally, 
\begin{equation}\label{CriTemp}\begin{array}{l}
T_c(\rho)=\sup\bigg[\hbox{$\beta^{-1}$ such that $\exists$ a family of fugacities $z(L ,\beta)$ with $\rho_{\Lambda_L,\beta}(z(L ,\beta))\to\rho$}\\
\hskip 2.3 cm\hbox{such that}\quad \onwl{\limsup}\limits_{r\to\infty}\onwl{\limsup}\limits_{L\to\infty}
\onwl{\sup}\limits_{\substack{x,y\in\Lambda_L:\,|x-y|\ge r}}
\gamma_{\Lambda_L,\beta}(x,y;z(L,\beta ))>0\bigg].\end{array}\end{equation}
This definition is appropriate for our purposes (without assuming any monotonicity of the interacting correlations with respect to temperature), as our main result provides an upper bound on the critical temperature.

It is well-known (see, e.g., \cite[Eq. (1.12)]{LSSY-05}) that at density $\rho>0$, the ideal-gas critical 
temperature $T_c^{(0)}(\rho)$ is given as
\begin{equation}\label{ideal-temperature}
 T_c^{(0)}(\rho)=4\pi\left(\frac{\rho}{\zeta(3/2)}\right)^{2/3}.
\end{equation}
We will consider the critical temperature shift under the presence of the interaction potential.

Our goal is to derive universal estimates which only depend on the potential via its scattering length. For the hard-sphere potential where $U(x)=+\infty$ for $|x|\le a$ 
and $U(x)=0$ otherwise, the scattering length is $a$. For a general nonnegative potential, the scattering length $a$ is defined by
\begin{equation}\label{scattering-definition}
 8\pi a=\inf_{\psi\in \cH^1(\RR^3;\RR)}\cE_U(\psi),\qquad
 \cE_U(\psi)=\int_{\RR^3}
       \bigl(2|\nabla\psi|^2+U|1-\psi|^2\bigr)\,dx.
\end{equation}
The factor 2 in  \eqref{scattering-definition} corresponds to the relative Hamiltonian $-2\Delta+U$, after  pinning 
the center of mass. For radial potentials this is the
usual convention in which the scattering solution $\psi$ is $a/|x|$ outside the support. We are interested in the dilute regime, with the dimensionless parameter $\xi =a\rho^{1/3} \ll 1$, where every particle is effectively felt by others as a hard-sphere with diameter $a$. Throughout, the interaction is fixed with $a>0$, and we  consider the low-density limit $\rho \to 0$.

\subsection{Main results}
\label{sec-main-comparison}

The interaction-induced shift in the critical temperature, $\Delta T_c(\rho)=T_c(\rho)-T_c^{(0)}(\rho)$, has been studied by various analytical and numerical methods. Early theoretical work includes the analysis of hard-sphere Bose gases by Lee and Yang~\cite{LY}. Later, Gr\"uter, Ceperley, and Lalo\"e~\cite{GCL} used path-integral Monte Carlo simulations to study the transition, while Holzmann and Krauth~\cite{HK} investigated the same problem using Monte Carlo methods. These studies provided important insights into the behavior of dilute homogeneous Bose gases near the transition in the thermodynamic limit. 

The mean-field approximation  predicts no shift in the critical temperature. The main difficulty in studying  $\Delta T_c(\rho)=T_c(\rho)-T_c^{(0)}(\rho)$ lies in the role of fluctuations beyond mean-field theory. In \cite[Eq.~(6)]{Huang99}, Huang used a first-order density formula and an optimization in the fugacity to predict that 
\begin{equation}\label{compc}
\dfrac{\Delta T_c(\rho)}{T_c^{(0)}(\rho)} \sim C_{\rm H}\sqrt{a\rho^{1/3}}, \quad  C_{\rm H}=\frac{2^{7/2}\sqrt\pi}{3\zeta(3/2)^{2/3}}
          =3.523992\ldots.
\end{equation}

In \cite[Sections~II--V]{SU}, Seiringer and Ueltschi combined a density lower
bound, Feynman cycles, and a scattering variational principle to prove rigorously an 
exponential decay of correlations above a threshold, which leads to the upper bound 
\begin{equation}\label{intro-SU-constant}
\frac{\Delta T_c(\rho)}{T_c^{(0)}(\rho)} \le  ( C_{\rm SU} + o(1)_{\rho \to 0}) \sqrt{a\rho^{1/3}},\quad 
 C_{\rm SU}=\frac{2^{7/2}\sqrt\pi}{3\zeta(3/2)^{7/6}}
                 \sqrt{2^{3/2}+\zeta(3/2)}
           =5.085675\ldots.
\end{equation}
           where $o(1)_{\rho\to 0}$ is a constant which tends to $0$ when $\rho \to 0$. 
          They also pointed out that a refinement of their density estimate would lead to Huang's coefficient. In the present paper, we establish  such a refinement and thereby improve their results. More precisely, we obtain the following main result. 

\begin{theorem}[Critical-temperature shift]
\label{theo-temperature}
Let $U$ be an even, nonnegative finite-range interaction
with scattering length $a>0$ in \eqref{scattering-definition}, and let $C_{\rm H}$ be the
Huang constant in \eqref{compc}.
Then, for every $\eta>0$, there exists $\rho_0>0$ such
that, whenever
\begin{equation}\label{temperature-bound}
0<\rho<\rho_0 \quad \text{and}\quad
\frac{T-T_c^{(0)}(\rho)}{T_c^{(0)}(\rho)}
>
(C_{\rm H}+\eta)\sqrt{a\rho^{1/3}},
\end{equation}
the following holds for all $L$ sufficiently large. Let $\Lam_L=[-L/2,L/2]^3$ and
$\beta=T^{-1}$. Consider any sequence of grand-canonical Gibbs states with fugacities $z=z(L,\beta)$ such that the mean
densities $\vrho_{\Lam_L,\beta}(z)$ converge to $\rho$
as $L\to\infty$. Then the kernel in \eqref{1prdm} satisfies an exponential off-diagonal estimate 
\begin{equation}
\gam_{\Lam_L,\beta}(x,y;z)\le B\exp\Big(-A|x-y|\Big),
\end{equation}
where constants $A,B>0$ are independent of $L$. 
Consequently, we have the critical-temperature shift upper bound 
\begin{equation}\label{critical-limsup}
	\limsup_{\rho\downarrow0}
	\frac{T_c(\rho)/T_c^{(0)}(\rho)-1}{\sqrt{a\rho^{1/3}}}
	\le C_{\rm H}.
\end{equation}
\end{theorem}

We note that Theorem \ref{theo-temperature} neither proves existence of a condensed phase nor determines the
sign or the true leading order of the temperature shift. The dilute limit is
$\rho\downarrow0$ with $U$ fixed; no uniformity for density-dependent families
of potentials is asserted. We use thermodynamic lower limits of the density,
so differentiability of the limiting pressure is not required.

Our result has the coefficient~\eqref{compc} from \cite[Eq.~(6)]{Huang99}, but only as
an \emph{upper bound}. The rigorous {\em lower bound} remains difficult since it is related to the proof of the off-diagonal long-range order at low temperature, which is a major question in mathematical physics.  

In fact, it is unclear if the full asymptotic formula \eqref{compc} should hold. In \cite{LY}, Lee and Yang used the pseudopotential method to analyze the excitation spectrum of the dilute Bose gas, and predicted  
a positive shift in the critical temperature of the {\em linear form}
\begin{equation}\label{intro-linear-prediction}
 \frac{\Delta T_c}{T_c^{(0)}}
       =c_{\rm lin}\,a\rho^{1/3}+o(a\rho^{1/3}).
\end{equation}
In a simplified model of hard spheres, they obtained $c_{\rm lin}\approx 1.79$ (this follows from the expansion of the critical density in \cite[Eq. (A2)]{LY}; see also \cite[Eq. (1.4)]{NRS}). 

Later, in \cite{Baym99,Baym01}, by considering finite-temperature Green's functions, Baym, Blaizot, Holzmann, Laloë, and Vautherin identified the infrared obstruction to an ordinary perturbative calculation and analyzed the contribution of critical, low-momentum fluctuations, also predicting a temperature shift similar to \eqref{intro-linear-prediction}. Numerical calculations for the associated three-dimensional classical field theory support this prediction, with $c_{\rm lin}=1.32\pm0.02$ in Arnold and Moore~\cite{AM} and $c_{\rm lin}=1.29\pm0.05$ in Kashurnikov, Prokof'ev, and Svistunov~\cite{KPS}. These are numerical determinations in the effective
theory, not rigorous bounds for the microscopic Hamiltonian~\eqref{HN}. Within the Bogoliubov free-energy functional, Napi\'orkowski, Reuvers and Solovej \cite{NRS} proved rigorously that  \eqref{intro-linear-prediction} holds with $c_{\rm lin} = 1.49$ in the limit when the first Born approximation approaches the scattering length  (see also \cite{NRSb} for related results in two dimensions). Our result neither establishes Huang's asymptotic
law \eqref{compc} nor contradicts a linear shift \eqref{intro-linear-prediction}.

Finally, we note that rigorous information is also available at the level of the free energy; see 
Seiringer~\cite{Seiringer08} and Yin~\cite{Yin10} for complementary lower
and upper bounds. However, such estimates do not justify
differentiating an asymptotic remainder with respect to the chemical
potential $\mu$ close to the transition. The approach below instead estimates the
grand-canonical density directly and keeps control as the fugacity $z$ approaches
1, which is crucial to obtain the bound in \eqref{critical-limsup}.

\subsection{The strategy}\label{sec-new-idea}

In this subsection we sketch the strategy of the proof of   Theorem \ref{theo-temperature}. We use the general
approach developed by Seiringer and Ueltschi \cite{SU}. Our main ingredient is Theorem~\ref{theo-density} below which gives
\begin{equation}\label{intro-density-bound}
{\vrholo_\beta(z)}\ge\lambda^{-3}
       \left[g_{3/2}(z)-\frac{4\aeff(\beta)}{\lambda}
                         g_{1/2}(z)g_{3/2}(z)\right],
       \qquad 0<z<1,
\end{equation}
where ${\vrholo_\beta(z)}$ is the thermodynamic lower limit along expanding
cubes and $\aeff(\beta)\to a$ as $\beta\to\infty$. 
{Here and below we use the free} heat kernel, thermal wavelength, and Bose functions
\begin{equation}\label{heat-notation}
 G_t(x)=(4\pi t)^{-3/2}e^{-|x|^2/(4t)},\qquad
 \lambda=\sqrt{4\pi\beta},\qquad
 g_s(z)=\sum_{k\ge1}\frac{z^k}{k^s}.
\end{equation}
The estimate {\eqref{intro-density-bound}} holds for
all $z\in (0,1)$, including choices $z=z(\beta)\uparrow 1$. It was explicitly observed in \cite{SU}, following their Theorem~III.1, that a density lower bound with the unshifted product $g_{1/2}(z)g_{3/2}(z)$ would give
Huang's coefficient. We retain the density-to-correlation argument  from \cite{SU} and
obtain the same structure with an effective scattering length.

Following the background from \cite{SU}, our bound \eqref{intro-density-bound} is a consequence of the error estimate in Proposition~\ref{prop-pair}. In fact, in the bound 
$\vrho_{\Lam,\beta}(z)\ge\vrho^{(0)}_{\Lam,\beta}(z)-\vrho_{\Lam,\beta}^A(z)-\vrho_{\Lam,\beta}^B(z)$ (cf. Proposition 
\ref{prop-decomposition}), the two error terms  $\vrho_{\Lam,\beta}^A(z)$ and $\vrho_{\Lam,\beta}^B(z)$ correspond with interactions
within one cycle and between two distinct cycles.

The main difficulty is to estimate these error terms efficiently. In principle, after removing the other interactions, each term involves a pair of marked Brownian segments. If the entire interaction interval of duration $\beta$ is retained,
the free connections have lengths $(m-1)\beta$ and $(n-1)\beta$, where
$m,n\ge1$. This removes a non-negligible amount of free propagation for
short windings, and a connecting interval disappears when $m=1$ or $n=1$.

Our  main improvement is to localize the interaction in time \emph{before} bounding the free heat kernels. To be precise, for an integer $M\ge2$, put $\tau=\beta/M$ and write the pair action as
$V=\sum_{r=0}^{M-1}V_r$, with $V_r$ the action on the $r$th interval.
By
positivity of the interaction potential and the elementary inequality $(1- a_1) \cdots (1- a_n) \ge 1 - a_1 -\cdots - a_n$ with $a_1,...,a_n \in [0,1]$, we have
\begin{equation}\label{intro-defect-splitting}
 1-e^{-V} = 1 - \prod_{r=0}^{M-1} (1- (1-e^{-V_r})) \le\sum_{r=0}^{M-1}(1-e^{-V_r}).
\end{equation}
Here no independence of the intervals is needed. Rerooting each free trace at
the beginning of the retained interval leaves connecting times
$m\beta-\tau$ and $n\beta-\tau$. Consequently,
\begin{equation}\label{intro-connector-bound}
 G_{m\beta-\tau}(0)G_{n\beta-\tau}(0)
       \le(1-M^{-1})^{-3}G_{m\beta}(0)G_{n\beta}(0),
       \qquad m,n\ge1.
\end{equation}
As we will see, we may choose $M\to \infty$ and at the same time keep $\beta/M\to \infty$ to retain the scattering length. 
Thus the loss in \eqref{intro-connector-bound} can be made small \emph{uniformly}, including adjacent legs
of a single loop and winding-one loops in the two-loop term. The two
connection patterns are kept distinct; they admit the same upper bound.

The above procedure leads to a slight change of the scattering length. To be precise, the scattering input remains the finite-time heat-kernel mass
$\mathfrak a(t)$ from~\cite[Section~V]{SU}; see also~\cite{LS}.
Here $K_t$ is the kernel of $e^{2t\Delta}-e^{t(2\Delta-U)}$.
In our normalization, $\iint K_t(x,y)\,dx\,dy=8\pi t\mathfrak a(t)$  where
$\mathfrak a(t)$ tends to the scattering length $a$ as $t\to \infty$. Thus the sum of the $M$ slice bounds contains
$M\tau=\beta$,  and no additional factor $M$. Combining this fact with
\eqref{intro-connector-bound} gives the coefficient
\[
 a_{\beta,M}=(1-M^{-1})^{-3}\mathfrak a(\tau),\quad \tau=\frac{\beta}{M}
\]
The intermediate scale is chosen so that
\begin{equation}\label{intro-two-scales}
 \tau\longrightarrow\infty,
       \qquad \frac{\tau}{\beta}\longrightarrow0.
\end{equation}
The first limit recovers the full scattering length, while the second
recovers the full free propagation times. In particular, the procedure is
not a short-time linearization in $U$ and remains valid for hard cores.

Note that the idea of splitting the time interval already appears in~\cite[Lemma~V.3]{SU}. The refinement here is to localize the interaction on a tunable intermediate time scale before estimating the free connecting heat kernels, while simultaneously retaining the scattering-length limit and uniform control of all winding numbers. This uniformity is important for the coefficient. The winding sum becomes
\begin{equation}\label{intro-winding-factorization}
 \sum_{m,n\ge1}m z^{m+n}G_{m\beta}(0)G_{n\beta}(0)
       =\lambda^{-6}g_{1/2}(z)g_{3/2}(z).
\end{equation}
As $z\uparrow1$, the factor  $g_{1/2}(z)$ diverges, while every fixed term in $g_{3/2}(z)$ contributes to its limiting value $\zeta(3/2)$. Keeping only long
windings in both variables would therefore lose part of the leading
coefficient. In~\cite[Eqs.~(13)--(14)]{SU}, the singular factor is instead
$(2^{3/2}+g_{3/2}(z))g_{1/2}(z)$. The estimate
\eqref{intro-density-bound} removes this extra $2^{3/2}$ contribution; the
resulting replacement in~\eqref{intro-SU-constant} is
\[
 \sqrt{2^{3/2}+\zeta(3/2)}\quad\longrightarrow\quad
 \sqrt{\zeta(3/2)},
\]
which gives the upper bound in~\eqref{compc}.

Finally, let us briefly explain how the density lower bound leads to our upper bound on the critical temperature.  As was said, we follow the density-to-correlation argument 
of \cite{SU}, but for completeness, we formulate it at finite volume. 
With $\alpha=-\log z>0$, the ideal density deficit is bounded by a
constant times $\sqrt\alpha$, whereas the interaction correction in
\eqref{intro-density-bound} is bounded by a constant times
$(\aeff/\lambda)/\sqrt\alpha$, after multiplication by $\lambda^3$.
Balancing these terms gives $\alpha$ of order $a/\lambda$ and a
square-root normal-phase bound.  Such an optimization gives an effective control in
absence of condensation; it does not identify the true critical point.

The rest of the paper is organized as follows.  In Section~\ref{sec-density}, we prove the density
bound, including the two-body normalization, the time-localization estimate,
and the separate winding counts for  $\vrho^A$ and $\vrho^B$.
Section~\ref{sec-temperature} proves the temperature bound at every
temperature above the threshold. Finally, Section~\ref{sec-hardcore} gives an explicit
hard-sphere error.

\bigskip
\noindent{\bf Acknowledgments.}  PTN was partially supported by the European Research Council via the ERC Consolidator
Grant RAMBAS Project-Nr. 10104424. TN's research was supported in part by the NSF grant DMS-2349981 and a Simons Foundation fellowship. YS’s research was partially supported by St John's College, Cambridge.

%%%%%%%%%%%%%%%%%%%%%
\section{Lower bounds on density}\label{sec-density}
%%%%%%%%%%%%%%%%%%%%%

Following \cite{SU}, we define the pointwise nonnegative two-body kernel difference $K_t(x,y)$ and its
normalized integral ${\mathfrak a}(t)$:
\begin{equation}\label{kernelK}
 K_t(x,y)=\bigl(e^{2t\Delta}-e^{t(2\Delta-U)}\bigr)(x,y),\qquad
 \mathfrak a(t)=\frac{1}{8\pi t}
       \int_{\RR^3}\int_{\RR^3}K_t(x,y)\,dx\,dy.
\end{equation}
Parameter ${\mathfrak a}(t)$ is the finite-time scattering quantity of~\cite[Section~V]{SU}; see
also~\cite{LS}. For every integer $M\ge2$, put
\begin{equation}\label{effective-definition}
 a_{\beta,M}=(1-M^{-1})^{-3}\mathfrak a(\beta/M),\qquad
 \aeff(\beta)=\inf_{M\ge2}a_{\beta,M}.
\end{equation}

\begin{theorem}[Density lower bound]\label{theo-density}
For every $\beta>0$, $0<z<1$, and bounded open $\Lambda\subset\RR^3$,
\begin{equation}\label{lowbound-finite}
{\vrho_{\Lambda,\beta}(z)
 \ge\vrho^{(0)}_{\Lambda,\beta}(z)}
   -\frac{\aeff(\beta)}{4\pi^2\beta^2}
                g_{1/2}(z)g_{3/2}(z),
\end{equation}
where {$\vrho^{(0)}_{\Lambda,\beta}(z)$} is the finite-volume ideal-gas density: 
\begin{equation}\label{free-density}
{\vrho^{(0)}_{\Lambda,\beta}(z)}
   =\frac{1}{|\Lambda|}\sum_{k\ge1}z^k
                       \Tr(e^{k\beta\Delta_\Lambda}).
\end{equation}
For cubes $\Lambda_L=[-L/2,L/2]^3$, define
$\vrholo_\beta(z)=\liminf_{L\to\infty}\vrho_{\Lambda_L,\beta}(z)$. Then
\begin{equation}\label{lowbound}
 \vrholo_\beta(z)\ge\lambda^{-3}
    \left[g_{3/2}(z)-\frac{4\aeff(\beta)}{\lambda}
                          g_{1/2}(z)g_{3/2}(z)\right].
\end{equation}
Moreover, the value $\aeff(\beta)$ obeys
\begin{equation}\label{effective-limit}
 a\le\aeff(\beta)<\infty,\qquad
 \lim_{\beta\to\infty}\aeff(\beta)=a
\end{equation}
with the scattering length $a$ in \eqref{scattering-definition}. The estimates hold simultaneously for all $0<z<1$, including fugacities $z(\beta)$ which approach $1$ as $\beta\to\infty$.
\end{theorem}

For the reader's convenience, we will quickly recall the Feynman--Kac representation and the density decomposition $
 \vrho_{\Lambda,\beta}(z)
     \ge\vrho^{(0)}_{\Lambda,\beta}(z)
                  -\vrho^A_{\Lambda,\beta}(z)-\vrho^B_{\Lambda,\beta}(z)$ from \cite{SU}. Then we explain how to bound the error terms  $\vrho^A_{\Lambda,\beta}(z)$ and $\vrho^B_{\Lambda,\beta}(z)$. 

\subsection{Feynman--Kac representation and density decomposition}\label{sec-fk}

The path-integral representation incorporates the Bose statistics through permutation
cycles, as in Feynman's approach~\cite{Feynman}. For a discussion of cycle
lengths and their relation to condensation, see~\cite{Ueltschi06}. We use
this representation to estimate the density, without assuming an equivalence
between infinite cycles and condensation. Following~\cite{SU},  let $dW^{t,\Lam}_{xy}$ 
denote the unnormalized Wiener-bridge measure with endpoints $x,y\in\Lam$, over the time-interval $[0,t]$, $t>0$, for the Laplacian
$\Delta$, restricted to paths in $\Lam$. The mass of this measure 
is the Dirichlet heat kernel $G_t^\Lambda(x,y)$. The full-space
measure is denoted by $dW^t_{xy}$, so that
\begin{equation}\label{bridge-mass}
 \int dW^t_{xy}=G_t(x-y)=\frac{e^{-|x-y|^2/(4t)}}{(4\pi t)^{3/2}},\qquad
 0\le G_t^\Lambda(x,y)\le G_t(x-y).
\end{equation}
For  $N\ge2$, the Feynman--Kac formula gives
\begin{equation}\label{expH}
 \Tr(e^{-\beta H_{\Lambda,N}})
 =\frac{1}{N!}\sum_{\sigma\in S_N}
   \int_{\Lambda^N}\!\int
   \prod_{i=1}^N dW^{\beta,\Lambda}_{x_i x_{\sigma(i)}}(\omega_i)
   e^{-\sum_{i<j\le N}\int_0^\beta U(\omega_i(s)-\omega_j(s))\,ds}
   \,dx_1\cdots dx_N.
\end{equation}
The bridge concatenation identity reads
\begin{equation}\label{dWprop}
 \int_\Lambda du \int dW^{t,\Lambda}_{xu}(\omega_1)
      dW^{s,\Lambda}_{uy}(\omega_2)
      F(\omega_1\mathbin{\star}\omega_2)
 =\int dW^{t+s,\Lambda}_{xy}(\omega)F(\omega).
\end{equation}
for any $t,s>0$, $x,y\in\Lam$ and a suitable (e.g., bounded and continuous)
function $F (\om)$. 
Here $u\in\Lambda$ is the intermediate point where the first path ends and the second begins. A similar identity holds for full-space measures 
$dW^t_{\cdot\;\cdot}$ and $dW^s_{\cdot\;\cdot}$ as well. Concatenating along the disjoint cycles of a permutation yields  that for $N\geq 2$
\begin{equation}\label{expH1}
 \Tr(e^{-\beta H_{\Lambda,N}})
 =\sum_{n=1}^N\frac{1}{n!}
   \sum_{\substack{k_1,\ldots,k_n\ge1\\k_1+\cdots+k_n=N}}
   \int\prod_{j=1}^n
   \left[\frac{dx_j}{k_j}
      dW^{k_j\beta,\Lambda}_{x_jx_j}(\omega_j)e^{-V(\omega_j)}\right]
   e^{-\sum_{i<j}V(\omega_i,\omega_j)}.
\end{equation}
Here a loop $\omega$ of winding number $k$ has duration $k\beta$, and its
legs are $\omega^{(i)}(s)=\omega((i-1)\beta+s)$ for $1\le i\le k$ and
$0\le s\le\beta$. We set
\begin{equation}\label{defVij}
 \begin{aligned}
 V_{ij}(\omega)&=\int_0^\beta
       U(\omega^{(i)}(s)-\omega^{(j)}(s))\,ds,
 &V(\omega)&=\sum_{1\le i<j\le k}V_{ij}(\omega),\\
 V_{ij}(\omega,\omega')&=\int_0^\beta
       U(\omega^{(i)}(s)-\omega'^{(j)}(s))\,ds,
 &V(\omega,\omega')&=\sum_{i=1}^k\sum_{j=1}^{k'}
                      V_{ij}(\omega,\omega').
 \end{aligned}
\end{equation}
Thus $V(\omega)$ accounts for interactions within a cycle, while
$V(\omega,\omega')$ accounts for interactions between distinct cycles.
On the disjoint union over winding numbers, introduce the measure
\begin{equation}\label{defmu}
 d\mu_\Lambda(\omega)
    =\frac{z^k}{k}\,dx\,dW^{k\beta,\Lambda}_{xx}(\omega)e^{-V(\omega)},
 \qquad x\in\Lambda.
\end{equation}
Integration against this measure includes summation over $k=k(\om)\ge1$ ($d\mu_\Lambda$ is not a probability measure). In the
Feynman--Kac representation, formulas \eqref{canZ}--\eqref{1prdm} take the 
following form 
\begin{equation}\label{canZ1}
\cZ_{\Lam ,\beta}(z)
 =1 + \int d\mu_\Lambda(\omega) {+\sum_{n\ge 2}}\frac{1}{n!}
     \int\prod_{j=1}^n d\mu_\Lam(\om_j)
       e^{-\sum_{1\le i<j}V(\om_i,\om_j)},
\end{equation}
\begin{equation}\label{density1}
\rho_{\Lam,\beta}(z)
=\dfrac{1}{|\Lam|\cZ_{\Lam ,\beta}(z)}\bigg[{\diy \int k(\omega)\,d\mu_\Lambda(\omega)} 
 +
    \sum\limits_{n\ge 2}\dfrac{1}{(n-1)!}
     \diy\int\prod\limits_{j=1}^n d\mu_\Lam(\om_j)k(\om_1)
            e^{-\sum_{1\le i<j}V(\om_i,\om_j)}\bigg] 
\end{equation}
and
\begin{equation}\label{1prdmFK}\begin{array}{l}
\gam_{\Lam,\beta}(x,y;z)=\dfrac{1}{\cZ_{\Lam,\beta}(z)}\bigg[
\sum\limits_{k\ge 1}z^k\int dW^{k\beta,\Lam}_{xy}(\om)e^{-V(\om)} +
\\
\qquad+\sum\limits_{n\ge 1}\dfrac{1}{n!}
   \diy\int\prod_{j=1}^n d\mu_\Lam(\om_j)e^{-\sum_{1\le i<j}V(\om_i,\om_j)}\diy\sum\limits_{k\ge 1}z^k\int dW^{k\beta,\Lam}_{xy}(\om)
e^{-V(\om)}e^{-\sum_{1\le l\le n}V(\om,\om_l)}\bigg].
\end{array}\end{equation}
Here $n$ counts the closed background loops, while the open path is treated 
separately. 

For singular potentials, including hard cores, the above
Feynman--Kac formulas are understood through the monotone
truncations $U_\ell=\min\{U,\ell\}$. The corresponding
finite-volume partition functions and density numerators
converge by dominated convergence for $0<z<1$, using the
ideal-gas expressions as dominating bounds. This justifies the singular and
hard-core interpretations throughout the paper.

In the following, we retain the decomposition of the density used in~\cite{SU}. For convenience,
write
\[
 \cT_{k,\Lambda}(F)
     =\int_\Lambda dx\int dW^{k\beta,\Lambda}_{xx}(\omega)F(\omega).
\]

\begin{proposition}[Density decomposition, \cite{SU}]\label{prop-decomposition}
One has
\begin{equation}\label{density-decomposition}
{\vrho_{\Lambda,\beta}(z)
     \ge\vrho^{(0)}_{\Lambda,\beta}(z)
                  -\vrho^A_{\Lambda,\beta}(z)-\vrho^B_{\Lambda,\beta}(z),}
\end{equation}
where
\begin{equation}\label{Adensity}
 \begin{aligned}
 {\vrho^A_{\Lambda,\beta}(z)}
   &=\frac{1}{|\Lambda|}\sum_{k\ge2}z^k\cT_{k,\Lambda}
       \left(\sum_{1\le i<j\le k}(1-e^{-V_{ij}(\omega)})\right),\\
 {\vrho^B_{\Lambda,\beta}(z)}
   &=\frac{1}{|\Lambda|}\iint k(\omega_1)^2k(\omega_2) 
       (1-e^{-V_{11}(\omega_1,\omega_2)})
                         d\mu_\Lambda(\omega_1)d\mu_\Lambda(\omega_2).
 \end{aligned}
\end{equation}
\end{proposition}

\begin{proof}
Isolate the first loop in \eqref{density1} and use
\[
 e^{-\sum_{j=2}^nV(\omega_1,\omega_j)}
   \ge1-\sum_{j=2}^n(1-e^{-V(\omega_1,\omega_j)}).
\]
The first term leaves the partition function of the remaining loops, which
cancels $\cZ_\Lambda$. In a negative term, dropping all interactions of the
marked second loop with the unmarked loops increases the magnitude of that
term. Summing over the unmarked loops again produces $\cZ_\Lambda$. Thus
\begin{equation}\label{marked-decomposition}
 \vrho_{\Lambda,\beta}(z)
 \ge\frac{1}{|\Lambda|}\int k(\omega)\,d\mu_\Lambda(\omega)
  -\frac{1}{|\Lambda|}\iint k(\omega_1)
         (1-e^{-V(\omega_1,\omega_2)})
                     d\mu_\Lambda(\omega_1)d\mu_\Lambda(\omega_2).
\end{equation}
Applying the bound $1-e^{-\sum_j v_j}\le\sum_j(1-e^{-v_j})$ to the self-interaction
in the first term gives  $\vrho^{(0)}-\vrho^A$.
Applying it to the mutual interaction in the second term gives $k(\omega_1)k(\omega_2)$
marked leg pairs. Each has the same integral: the self-interacting measure
of either loop is invariant under shifts by integer multiples of $\beta$.
Together with the factor $k(\omega_1)$ in \eqref{marked-decomposition}, this gives
the factor $k(\omega_1)^2k(\omega_2)$ in \eqref{Adensity}.
These rearrangements concern nonnegative integrands, so Tonelli's theorem
applies. The error terms are finite by the estimates proved below. The
argument extends to singular interactions by monotone approximation.
\end{proof}

\subsection{The two-body kernel and finite-time scattering}\label{sec-kernel}

For $t>0$, let
\begin{equation}\label{defect-definition}
 \begin{split}
 \cD_t(x_1,x_2;x_3,x_4)
 ={}&\int dW^t_{x_1x_2}(\omega_1)dW^t_{x_3x_4}(\omega_2)\left(1-e^{-\int_0^t
           U(\omega_1(s)-\omega_2(s))\,ds}\right).
 \end{split}
\end{equation}
Its finite-volume analogue is denoted by $\cD_t^\Lambda$. We have  $0\le\cD_t^\Lambda\le\cD_t$ pointwise. 

\begin{lemma}[Two-body normalization]\label{lem-kernel}
With the heat kernel in \eqref{heat-notation},
\begin{equation}\label{Winteract}
 \cD_t(x_1,x_2;x_3,x_4)
   =8G_{2t}(x_1-x_2+x_3-x_4)
                         K_t(x_1-x_3,x_2-x_4).
\end{equation}
Furthermore, for  every fixed $x_4\in\RR^3$,
\begin{equation}\label{integrated-mass}
 \int_{\RR^9}\cD_t(x_1,x_2;x_3,x_4)\,dx_1dx_2dx_3
       =8\pi t\,\mathfrak a(t).
\end{equation}
\end{lemma}

\begin{proof}
For two independent paths with generator $\Delta$, the center of mass
$c=(\omega_1+\omega_2)/2$ has generator $\Delta/2$ and the relative
coordinate $r=\omega_1-\omega_2$ has generator $2\Delta$. The corresponding
Gaussian processes are independent, and the coordinate change has Jacobian
one. The center-of-mass kernel is
\[
 G_{t/2}\!\left(\frac{x_1+x_3-x_2-x_4}{2}\right)
   =8G_{2t}(x_1-x_2+x_3-x_4),
\]
while the relative interaction defect is $K_t$. This proves \eqref{Winteract}.
For the integrated identity, set $r=x_1-x_3$, $r'=x_2-x_4$, and
$b=x_3-x_4$. The Jacobian is again one, and
\[
 \int_{\RR^3}8G_{2t}(r-r'+2b)\,db=1.
\]
The remaining integral is $\diy{\iint}K_t(r,r')\,dr\,dr'$, which proves
\eqref{integrated-mass}.
\end{proof}

\begin{lemma}[Finite-time scattering estimate]\label{lem-scattering}
For every $\psi\in \cH^1(\RR^3;\RR)$ and $t>0$,
\begin{equation}\label{finite-time-bound}
 a\le\mathfrak a(t)
     \le\frac{\cE_U(\psi)}{8\pi}+\frac{\|\psi\|_2^2}{4\pi t}.
\end{equation}
In particular, $\mathfrak a(t)$ is finite and tends to $a$ as $t\to\infty$.
\end{lemma}

\begin{proof}

First assume $U$ is bounded. We recall the variational identity of~\cite[Lemma~V.1]{SU}: 
\begin{equation}\label{finite-time-variational}
 8\pi\mathfrak a(t)
   =\inf_{\psi\in \cH^1(\RR^3;\RR)}
     \left\{\cE_U(\psi)+t^{-1}\langle\psi,F(tH)\psi\rangle\right\}, \quad F(s)=\frac{s(1-e^{-s})}{s-1+e^{-s}}, 
\end{equation}
where $H=-2\Delta+U>0$ on $L^2(\mathbb{R}^3)$.  The elementary inequalities $1\le F(s)\le2$ for $s>0$ imply
\eqref{finite-time-bound}. 

For unbounded $U$, apply the bound to $U_\ell=\min\{U,\ell\}$. The kernel differences
increase to $K_t$, and the scattering lengths increase to $a$; see \cite[Lemma 3.2]{BFS} for the latter assertion, including hard
cores.
 Thus for any fixed admissible $\psi$, passage to the limit $\ell\to \infty$ gives
\eqref{finite-time-bound}. Then optimizing over $\psi$ we find that $\mathfrak a(t)\to a$  as $t\to\infty$.
\end{proof}

\subsection{Time localization and the pair estimate}\label{sec-pair}

Let $m,n\ge1$. Define the integrated defect for a single loop by
\begin{equation}\label{pairA-definition}
 I^A_{m,n}=\frac{1}{|\Lambda|}\cT_{m+n,\Lambda}
                        (1-e^{-V_{1,m+1}(\omega)}),
\end{equation}
and the integrated defect for two distinct loops by
\begin{equation}\label{pairB-definition}
 I^B_{m,n}=\frac{1}{|\Lambda|}
        (\cT_{m,\Lambda}\otimes\cT_{n,\Lambda})
                        (1-e^{-V_{11}(\omega_1,\omega_2)}).
\end{equation}
These are free trace measures with one retained interaction defect; they
contain neither self-interaction weights nor powers of $z$. Dependence on
$\Lambda$ and $\beta$ is suppressed in this notation.

\begin{proposition}[Pair estimate]\label{prop-pair}
For every integer $M\ge2$ and all $m,n\ge1$,
\begin{equation}\label{pair-bound}
 I^A_{m,n},\ I^B_{m,n}
     \le8\pi\beta\,a_{\beta,M}G_{m\beta}(0)G_{n\beta}(0),
\end{equation}
where $a_{\beta,M}$ is defined in \eqref{effective-definition}. The estimate is uniform 
in $m,n$ and in the volume.
\end{proposition}

\begin{proof}
Set $\tau=\beta/M$. For a pair of marked legs, positivity gives the 
inequality
\begin{equation}\label{time-slicing}
 1-e^{-\int_0^\beta U(\omega_1(s)-\omega_2(s))\,ds}
 \le\sum_{l=0}^{M-1}
     \left(1-e^{-\int_{l\tau}^{(l+1)\tau}
                         U(\omega_1(s)-\omega_2(s))\,ds}\right).
\end{equation}
No independence of the slices and no smallness of $U$ are assumed.

The free trace measure of duration $t$ is invariant under every cyclic time
shift $\omega(s)\mapsto\omega(s+u\pmod t)$. For a cylinder functional, write
its integral as a cyclic product of Dirichlet heat kernels and integrate
out the root using the semigroup property. The resulting expression depends
only on the cyclic time differences. A monotone-class argument extends this
identity to measurable nonnegative functionals. Thus we can reroot at the
beginning of each retained interval. We use this continuous rerooting only
for the free trace measures in \eqref{pairA-definition}--\eqref{pairB-definition}.

For one slice in the single-loop geometry, the two interacting segments
have length $\tau$ and start $m\beta$ apart. Their connecting free segments
therefore have durations
\begin{equation}\label{connecting-times}
 m\beta-\tau,\qquad n\beta-\tau.
\end{equation}
Both are positive, even when $m=1$ or $n=1$. Concatenation gives the slice
integral
\begin{equation}\label{sliceA}
 \frac{1}{|\Lambda|}\int_{\Lambda^4}
 G^\Lambda_{m\beta-\tau}(x_2,x_3)
 G^\Lambda_{n\beta-\tau}(x_4,x_1)
 \cD^\Lambda_\tau(x_1,x_2;x_3,x_4)\,dx_1dx_2dx_3dx_4.
\end{equation}
For two distinct loops, reroot both at the beginning of the retained slice.
Their free remainders again have the durations in \eqref{connecting-times},
but the connections are different:
\begin{equation}\label{sliceB}
 \frac{1}{|\Lambda|}\int_{\Lambda^4}
 G^\Lambda_{m\beta-\tau}(x_2,x_1)
 G^\Lambda_{n\beta-\tau}(x_4,x_3)
 \cD^\Lambda_\tau(x_1,x_2;x_3,x_4)\,dx_1dx_2dx_3dx_4.
\end{equation}
In each of integrals \eqref{sliceA} and \eqref{sliceB}, we bound the connecting kernels
by $G_{m\beta-\tau}(0)$ and $G_{n\beta-\tau}(0)$, replace
$\cD^\Lambda_\tau$ by $\cD_\tau$, extend the $x_1,x_2,x_3$ integrals to
$\RR^3$, and keep $x_4\in\Lambda$. By \eqref{integrated-mass}, each slice
is bounded above by
\begin{equation}\label{one-slice-bound}
 8\pi\tau\,\mathfrak a(\tau)
                     G_{m\beta-\tau}(0)G_{n\beta-\tau}(0).
\end{equation}
For $m,n\ge1$,
\begin{equation}\label{heat-time-loss}
 G_{m\beta-\tau}(0)G_{n\beta-\tau}(0)
 \le(1-M^{-1})^{-3}G_{m\beta}(0)G_{n\beta}(0).
\end{equation}
All $M$ slices satisfy the same estimate. Summing over \eqref{time-slicing}
and using $M\tau=\beta$ proves \eqref{pair-bound}.
\end{proof}

\begin{remark}
The order of operations in the proof is essential: the interaction is
localized \emph{before} the free kernels are bounded. The factor $M$ from
summing over the slices cancels against their duration $\tau$. The retained
interval can be long in absolute time, while occupying a vanishing fraction
of a thermal leg.
	\end{remark}

\subsection{Upper bounds on \texorpdfstring{$\vrho^A(z)$}{rho A} and \texorpdfstring{$\vrho^B(z)$}{rho B}}

\begin{proposition}\label{prop-rhoA}
For every $M\ge2$ and $0<z<1$,
\begin{equation}\label{uprhoA}
 {\vrho^A_{\Lambda,\beta}(z)}
    \le8\pi\beta\,a_{\beta,M}\lambda^{-6}
                            g_{1/2}(z)g_{3/2}(z).
\end{equation}
\end{proposition}

\begin{proof}
Put $q_m=G_{m\beta}(0)=\lambda^{-3}m^{-3/2}$. Since $U$ is even, the sum
over unordered pairs of legs is half the sum over ordered distinct pairs.
Cyclic invariance then gives the exact counting identity
\begin{equation}\label{bdrhoA1}
 {\vrho^A_{\Lambda,\beta}(z)}
    =\frac12\sum_{k\ge2}kz^k\sum_{m=1}^{k-1}I^A_{m,k-m}.
\end{equation}
This includes $k=2$, for which there is one unordered pair. By
Proposition~\ref{prop-pair}, we bound
\begin{align*}
 {\vrho^A_{\Lambda,\beta}(z)}
 &\le4\pi\beta a_{\beta,M}
         \sum_{m,n\ge1}(m+n)z^{m+n}q_mq_n\\
 &=8\pi\beta a_{\beta,M}
       \left(\sum_{m\ge1}m z^mq_m\right)
       \left(\sum_{n\ge1}z^nq_n\right)=8\pi\beta a_{\beta,M}\lambda^{-6}
                                 g_{1/2}(z)g_{3/2}(z). 
\end{align*}
\end{proof}

\begin{proposition}\label{prop-rhoB}
For every $M\ge2$ and $0<z<1$,
\begin{equation}\label{uprhoB}
 {\vrho^B_{\Lambda,\beta}(z)}
    \le8\pi\beta\,a_{\beta,M}\lambda^{-6}
                            g_{1/2}(z)g_{3/2}(z).
\end{equation}
\end{proposition}

\begin{proof} 
Drop the two self-interaction factors in $d\mu_\Lambda$ in
\eqref{Adensity}. This is an upper bound, not an equality. If the winding
numbers are $m,n$, the cycle weights and marked-leg factors give
\[
 \frac{z^m}{m}\frac{z^n}{n}\,m^2n=m z^{m+n}.
\]
Consequently, Proposition~\ref{prop-pair} implies
\begin{align*}
 {\vrho^B_{\Lambda,\beta}(z)}
  &\le\sum_{m,n\ge1}m z^{m+n}I^B_{m,n}\\
  &\le8\pi\beta a_{\beta,M}
           \sum_{m,n\ge1}m z^{m+n}q_mq_n=8\pi\beta a_{\beta,M}\lambda^{-6}g_{1/2}(z)g_{3/2}(z).
\end{align*}
\end{proof}

\begin{remark}
Although the two error terms  $\vrho^A_{\Lambda,\beta}(z)$ and $\vrho^B_{\Lambda,\beta}(z)$ arise from different loop geometries, they satisfy the same upper bound thanks to  Proposition~\ref{prop-pair}. 
\end{remark}

\subsection{Proof of Theorem~\ref{theo-density}}

We are ready to complete the proof of Theorem~\ref{theo-density}. Indeed, combining \eqref{density-decomposition}, \eqref{uprhoA}, and \eqref{uprhoB},
we obtain, for every $M\ge2$,
\begin{equation}\label{density-with-M}
 \vrho_{\Lambda,\beta}(z)
  \ge \vrho^{(0)}_{\Lambda,\beta}(z)
       -16\pi\beta a_{\beta,M}\lambda^{-6}
                         g_{1/2}(z)g_{3/2}(z).
\end{equation}
Since $16\pi\beta\lambda^{-6}=1/(4\pi^2\beta^2)$, taking the infimum over
$M$ proves \eqref{lowbound-finite}. Attainment of the infimum is unnecessary. For a cube of side $L$ and each fixed $t>0$,
\[
 \frac{1}{L^3}\Tr(e^{t\Delta_{\Lambda_L}})
 =\left(\frac1L\sum_{j\ge1}e^{-\pi^2 t j^2/L^2}\right)^3
 \longrightarrow(4\pi t)^{-3/2}.
\]
The left side is at most $(4\pi t)^{-3/2}$ by comparison with the integral
of the decreasing Gaussian. Dominated convergence in \eqref{free-density}
therefore gives
$$
 \vrho^{(0)}_{\Lambda_L,\beta}(z) \longrightarrow \lambda^{-3}g_{3/2}(z).
$$
Taking a lower limit in \eqref{lowbound-finite} proves \eqref{lowbound}.

Lemma~\ref{lem-scattering} implies $\aeff(\beta)\ge a$ and finiteness.
For any fixed integer $M\ge2$,
$$
 \limsup_{\beta\to\infty}\aeff(\beta)
       \le(1-M^{-1})^{-3}a.
$$
Letting $M\to\infty$ proves \eqref{effective-limit}. This two-step limit
avoids any unstated uniformity in the scattering estimate.
This ends the proof of Theorem~\ref{theo-density}. 

%%%%%%%%%%%%%%%%%%%%%
\section{The critical-temperature bound}\label{sec-temperature}
%%%%%%%%%%%%%%%%%%%%%

\subsection{A finite-volume normal-phase criterion}

Let $\gamma_{\Lambda,\beta}(x,y;z)$ be as in \eqref{1prdm}. 
The Feynman--Kac representation with one open path, with all its nonnegative
interactions removed, gives the comparison used in~\cite[Section~II]{SU}:
\begin{equation}\label{gamma-bound}
 0\le \gamma_{\Lambda,\beta}(x,y;z) 
       \le\sum_{k\ge1}z^kG_{k\beta}(x-y),\qquad 0<z<1.
\end{equation}
The closed background loops contribute the partition function, which
cancels the normalizing factor. If $\alpha=-\log z>0$, then
\[
 \alpha k+\frac{|x-y|^2}{4\beta k}
    \ge\frac{\alpha k}{2}+|x-y|\sqrt{\frac{\alpha}{2\beta}},
\]
and hence
\begin{equation}\label{gamma-exponential}
\gamma_{\Lambda,\beta}(x,y;z)
 \le\lambda^{-3}g_{3/2}(e^{-\alpha/2})
        e^{-|x-y|\sqrt{\alpha/(2\beta)}}.
\end{equation}
Let $\gam_{\Lam ,\beta}(z)$ be the operator in $\rL^2(\Lam )$ with 
integral kernel $\gamma_{\Lambda,\beta}(x,y;z)$. Integrating the right side of \eqref{gamma-bound} in either variable and using Schur's test yields the following bound  for the norm 
$\|\gam_{\Lambda,\beta}(z)\|$:
\begin{equation}\label{gamma-norm}
\|\gamma_{\Lambda,\beta}(z)\|\le\frac{z}{1-z}.
\end{equation}

Suppose that
\begin{equation}\label{normal-density-criterion}
 \rho<\vrholo_\beta(z_*)\qquad\hbox{for some }z_*<1.
\end{equation}
For sufficiently large cubes, $\rho< \vrho_{\Lambda_L,\beta}(z_*)$ with a
strict gap uniform in $L$. The finite-volume density is continuous and
strictly increasing for $0<z<1$. In fact, 
\begin{equation}\label{variance}
 z\dfrac{\partial}{\partial z} \vrho_{\Lambda,\beta}(z)
       =\dfrac{1}{|\Lambda|}\operatorname{Var}_{\Lambda,\beta,z}[N].\end{equation}
Here $\operatorname{Var}_{\Lambda,\beta,z}[N]$ stands for the particle number variance
${\mathbb E}_{\Lambda,\beta,z}[N^2]-({\mathbb E}_{\Lambda,\beta,z}[N])^2$ in the 
grand-canonical Gibbs state:
$$\begin{array}{l}{\rm{Var}}_{\Lambda,\beta,z}[N]=\dfrac{1}{\cZ_{\Lam,\beta}(z)}
\sum\limits_{N\ge 1}N^2z^N\Tr(e^{-\beta H_{\Lambda,N}})-\left(\dfrac{1}{\cZ_{\Lam,\beta}(z)}
\sum\limits_{N\ge 1}Nz^N\Tr(e^{-\beta H_{\Lambda,N}})\right)^2.\end{array}$$
As $N$ is non-constant, we have 
$\operatorname{Var}_{\Lambda,\beta,z}[N]>0$. Consequently, $\vrho_{\Lambda,\beta}(z)$ 
is strictly increasing in $z$ throughout its domain of finiteness. Since 
$\vrho_{\Lambda,\beta}(z)\downarrow 0$ as $z\downarrow0$, there is a fugacity $z_L<z_*$
with mean density exactly $\rho$. More generally, any sequence of
well-defined grand-canonical Gibbs states whose mean densities tend to $\rho$
must have $z_L<z_*$ eventually; otherwise monotonicity contradicts the strict
gap. Equations \eqref{gamma-exponential} and \eqref{gamma-norm}, evaluated
at $z_*$, then hold uniformly for these states. This proves the normal-phase
criterion \eqref{normal-density-criterion}, without assuming a differentiable
thermodynamic pressure or an equivalence-of-ensembles theorem.

\subsection{Choice of fugacity and Huang's coefficient}

Put $Z_0=\zeta(3/2)$, $h=\aeff(\beta)/\lambda$, and $z=e^{-\alpha}$.
The decreasing-integrand comparison gives
\begin{equation}\label{g-half}
 g_{1/2}(e^{-\alpha})
       \le\int_0^\infty e^{-\alpha t}t^{-1/2}\,dt
       =\sqrt{\pi/\alpha}.
\end{equation}
Integrating with respect to $\alpha$ gives
\begin{equation}\label{g-three-half}
 Z_0-g_{3/2}(e^{-\alpha})
       =\int_0^\alpha g_{1/2}(e^{-s})\,ds
       \le2\sqrt{\pi\alpha}.
\end{equation}
Using also $g_{3/2}(z)\le Z_0$ in \eqref{lowbound}, we obtain
\begin{equation}\label{optimize}
 \lambda^3 \vrholo_\beta (e^{-\alpha})
       \ge Z_0-2\sqrt\pi\sqrt\alpha
                    -4hZ_0\sqrt\pi\,\alpha^{-1/2}.
\end{equation}
The right side is largest at
\begin{equation}\label{optimal-fugacity}
 \alpha_*=2Z_0h,\qquad
 z_* =\exp\left(-\frac{2Z_0\aeff(\beta)}{\lambda}\right)<1,
\end{equation}
and consequently
\begin{equation}\label{optimized-density}
 \vrholo_\beta(z_*)
       \ge\lambda^{-3}\left[
       Z_0-4\sqrt{2\pi Z_0}\sqrt{\frac{\aeff(\beta)}{\lambda}}\right].
\end{equation}
All these statements are inequalities, not asymptotic expansions in $z$.

Let $\xi=a\rho^{1/3}$ and write $T=T_c^{(0)}(\rho)(1+\theta)$  where $T_c^{(0)}(\rho)$
is the critical temperature of the free Bose-gas. Then
\begin{equation}\label{temperature-rescaling}
 \lambda^3\rho=Z_0(1+\theta)^{-3/2},\qquad
 \frac{\aeff(\beta)}{\lambda}
       =\frac{\aeff(\beta)}a\frac{\xi}{Z_0^{1/3}}(1+\theta)^{1/2}.
\end{equation}
The sufficient condition \eqref{normal-density-criterion} follows whenever
\begin{equation}\label{exact-temperature-criterion}
 1-(1+\theta)^{-3/2}
 >4\sqrt{2\pi}\,Z_0^{-2/3}
       \sqrt{\frac{\aeff(\beta)}a}\sqrt{\xi}\,(1+\theta)^{1/4}.
\end{equation}
In a bounded critical window, $\beta\to\infty$ as $\rho\downarrow0$, and
$\aeff(\beta)/a\to1$. If $\theta=O(\sqrt{\xi})$, then
\[
 1-(1+\theta)^{-3/2}=\frac32\theta+O(\theta^2),\qquad
 (1+\theta)^{1/4}=1+O(\theta).
\]
Thus \eqref{exact-temperature-criterion} holds at
$\theta=(C_{\rm H}+\eta)\sqrt{\xi}$, for small $\xi$ and any fixed $\eta>0$,
since
\begin{equation}\label{CH-calculation}
 \frac32 C_{\rm H}=4\sqrt{2\pi}\,Z_0^{-2/3}.
\end{equation}
This computes the coefficient. The next step extends the conclusion to all
larger temperatures; no monotonicity of interacting correlations in $T$ is
assumed.

\subsection{Uniformity at all higher temperatures}

\begin{proof}[Proof of Theorem~\ref{theo-temperature}]
Fix $\eta>0$, and set
\[
 \beta_* =\frac{a^2}{\xi},\qquad T_* =\beta_*^{-1},\qquad
 T_{\rm b}=T_c^{(0)}(\rho)[1+(C_{\rm H}+\eta)\sqrt{\xi}].
\]
For small $\xi$, $T_{\rm b}<T_*$. By \eqref{effective-limit},
\[
 \delta_\xi=\sup_{\beta\ge\beta_*}\left(\frac{\aeff(\beta)}a-1\right)
       \longrightarrow0.
\]
For $0<T\le T_*$, the right side of \eqref{optimized-density} is at least
\begin{equation}\label{monotone-lower}
 F_\xi(T)=\left(\frac{T}{4\pi}\right)^{3/2}
       \left[Z_0-4\sqrt{2\pi Z_0}\sqrt{a(1+\delta_\xi)}
                         \left(\frac{T}{4\pi}\right)^{1/4}\right].
\end{equation}
The sign of $F_\xi'(T)$ is the sign of
\[
 \frac32Z_0-7\sqrt{2\pi Z_0}\sqrt{a(1+\delta_\xi)}
                         \left(\frac{T}{4\pi}\right)^{1/4},
\]
which is positive throughout $0<T\le T_*$ for small $\xi$, since the second
term is $O(\xi^{1/4})$. The preceding critical-window calculation, with
$1+\delta_\xi$ in place of $\aeff/a$, gives $F_\xi(T_{\rm b})>\rho$ for small
$\xi$. Therefore \eqref{normal-density-criterion} holds for every
$T\in[T_{\rm b},T_*]$.

For $T\ge T_*$, equivalently $0<\beta\le\beta_*$, use
\eqref{density-with-M} with $M=2$ and choose $z=2\rho\lambda^3$.
Uniformly in this range,
\[
 z\le2(4\pi)^{3/2}\xi^{3/2}<\tfrac12
\]
for small $\xi$. The bounds $g_{3/2}(z)\ge z$ and
$g_{1/2}(z),g_{3/2}(z)\le z/(1-z)\le2z$ give
\begin{equation}\label{high-temperature-density}
 \vrholo_\beta(z)\ge2\rho-256\pi\rho^2\beta a_{\beta,2}.
\end{equation}
Choose a fixed compactly supported $\psi\in \cH^1(\RR^3;\RR)$ equal to one on
$B_{R_0}$, and put $A=\cE_U(\psi)/(8\pi)$ and $B=\|\psi\|_2^2/(4\pi)$.
By Lemma~\ref{lem-scattering},
\[
 \beta a_{\beta,2}=8\beta\mathfrak a(\beta/2)
                       \le8A\beta+16B.
\]
Consequently,
\begin{equation}\label{high-temperature-smallness}
 \sup_{0<\beta\le\beta_*}\rho\beta a_{\beta,2}
       \le\frac{8A}{a}\xi^2+\frac{16B}{a^3}\xi^3\longrightarrow0.
\end{equation}
The right side of \eqref{high-temperature-density} is therefore strictly
larger than $\rho$, uniformly for $\beta\le\beta_*$. This proves the
normal-phase criterion for all $T\ge T_*$ as well. In view of the definition of the critical temperature in \eqref{CriTemp}, the proof of  Theorem~\ref{theo-temperature} is complete. 
\end{proof}

%%%%%%%%%%%%%%%%%%%%%
\section{Hard-sphere interactions and an explicit error}\label{sec-hardcore}
%%%%%%%%%%%%%%%%%%%%%

For the  hard-sphere potential $U(x)=+\infty$ for $|x|\le a$ and $U(x)=0$ otherwise, 
the finite-time quantity ${\mathfrak a}(t)$ can be evaluated explicitly.

\begin{proposition}\label{prop-hardcore}
For a hard core of radius $a$ and every $t>0$,
\begin{equation}\label{hardcore-mass}
 {\mathfrak a}(t)
       =a+\sqrt{\frac2\pi}\frac{a^2}{\sqrt t}+\frac{a^3}{6t}.
\end{equation}
It follows that
\begin{equation}\label{hardcore-effective}
 \frac{a_{\beta,M}}a
  =(1-M^{-1})^{-3}
       \left[1+\sqrt{\frac2\pi}\,a\sqrt{\frac M\beta}
                                      +\frac{a^2M}{6\beta}\right]
\end{equation}
and, as $\beta/a^2\to\infty$,
\begin{equation}\label{hardcore-error}
 1\le\frac{\aeff(\beta)}a
       \le1+O\left((a^2/\beta)^{1/3}\right).
\end{equation}
\end{proposition}

\begin{proof}
Let $X_t$ be Brownian motion with generator $2\Delta$, and let $\tau_a$ be
its first entrance time into the ball of radius $a$, with $\tau_a=0$ for
starting points in the ball. Integrating the killed kernel gives
\begin{equation}\label{hardcore-hitting}
 8\pi t\,\mathfrak a_{\rm hs}(t)
       =\int_{\RR^3}\PP_y(\tau_a\le t)\,dy.
\end{equation}
For $r>a$, the bounded radial solution of the heat equation with diffusivity
two, boundary value one at $r=a$, and initial value zero is
\[
 \PP_r(\tau_a\le t)
       =\frac ar\operatorname{erfc}\left(\frac{r-a}{\sqrt{8t}}\right),
 \qquad
 \operatorname{erfc}(s)=\frac2{\sqrt\pi}\int_s^\infty e^{-u^2}\,du.
\]
This follows by multiplying the radial equation by $r$ and solving the
one-dimensional half-line heat equation. Including the volume of the core,
\begin{align*}
 8\pi t\,\mathfrak a_{\rm hs}(t)
  &=\frac{4\pi a^3}{3}
      +4\pi a\int_0^\infty(a+u)
                    \operatorname{erfc}\left(\frac{u}{\sqrt{8t}}\right)du\\
  &=8\pi at+8\sqrt{2\pi}\,a^2\sqrt t+\frac{4\pi a^3}{3}.
\end{align*}
This proves \eqref{hardcore-mass} and \eqref{hardcore-effective}.
Set $b=a^2/\beta$ and choose $M=\max\{2,\lceil b^{-1/3}\rceil\}$.
Then $M^{-1}=O(b^{1/3})$, $\sqrt{bM}=O(b^{1/3})$, and
$bM=O(b^{2/3})$, which proves \eqref{hardcore-error}.
\end{proof}

In the critical window we have $a^2/\beta=O(\xi^2)$, where $\xi=a\rho^{1/3}$.
Thus the relative scattering correction in
\eqref{exact-temperature-criterion} is $O(\xi^{2/3})$ for hard spheres.
The error in the resulting temperature criterion is controlled without
replacing the hard core by an integrable weak potential.

\end{document}